\documentclass[prl,twocolumn,superscriptaddress,showpacs,floatfix]{revtex4}
\usepackage{graphicx}
\usepackage{amssymb}
\usepackage{amsmath}
\usepackage{xspace}
\usepackage{url}

\begin{document}

\title{Quantum phase transition in a 1D transport model with 
boson affected hopping: Luttinger liquid versus 
charge-density-wave behavior}
\author{S. Ejima}
\affiliation{
Institut f\"ur Physik, Ernst-Moritz-Arndt-Universit{\"a}t
Greifswald, 17489 Greifswald, Germany }
\author{G. Hager}
\affiliation{
Regionales Rechenzentrum Erlangen, Universit\"at Erlangen-N\"urnberg,  
91058 Erlangen, Germany }
\author{H.~Fehske}
\affiliation{
Institut f\"ur Physik, Ernst-Moritz-Arndt-Universit{\"a}t Greifswald, 
17489 Greifswald, Germany }
\begin{abstract}
We solve a very general two-channel fermion-boson model 
describing charge transport within some background medium 
by means of a refined pseudo-site density matrix renormalization group 
(DMRG) technique.  Performing a careful finite-size scaling analysis, 
we determine the ground-state phase diagram and convincingly prove 
that the model exhibits a metal-insulator quantum phase transition
for the half-filled band case. In order to characterize the 
metallic and insulating regimes we calculate besides the local 
particle densities and fermion-boson correlation functions,
the kinetic energy, the charge structure factor, the Luttinger 
liquid charge exponent and the single-particle excitation gap 
for a one-dimensional infinite system.   
\end{abstract}
\maketitle
The proof of existence of metal-insulator transitions (MITs) in 
generic model Hamiltonians is one of the most fundamental problems in solid 
state theory. While the mechanisms that can drive 
a MIT, such as band structure effects~\cite{Bl29}, disorder~\cite{An58}, 
Coulomb correlations~\cite{Mo90}, or the coupling to the lattice 
degrees of freedom~\cite{Pe55}, are accepted in general, there 
is only a very small number of microscopic models which  
have rigorously been shown to indeed exhibit such a transition. 
Examples are the 3D Anderson tight-binding (disorder) model for which  
an analytical proof of particle localization 
exists~\cite{FMSS85}, 
or the 1D spinless fermion Holstein (electron-phonon) model, where the 
Tomonaga-Luttinger-liquid (TLL) charge-density-wave (CDW) MIT 
has been confirmed numerically by DMRG~\cite{BMH98}. 
Zero-temperature MITs triggered 
by Coulomb interaction are more difficult to assess; the prototype 
half-filled 1D Hubbard model, e.g., is insulating for all 
$U>0$~\cite{EFGKK05}, and only on introducing a (particular) 
long-range hopping the MIT takes place at finite 
interaction strength~\cite{GR92}. In the 1D half-filled extended Hubbard 
($U$-$V$) model, there exists at most a metallic line at the
bond-order- charge-density-wave insulator-insulator transition~\cite{EN07}. 
If Coulomb and electron-phonon interactions compete, an extended 
intervening metallic phase may occur between Mott and Peierls
insulating states, which allows for a MIT\@. This has been demonstrated  
for the 1D Holstein-Hubbard model at half filling~\cite{TAA05}.  

Quite recently a novel quantum transport Hamiltonian has been 
proposed~\cite{Ed06}, which describes regimes of 
quasi-free, correlation or fluctuation dominated transport. 
In a sense this model parameterizes the correlations inherent to 
a fermionic many-particle system, but also the couplings to 
phonon or bath degrees of freedom, by a ``background medium'' 
that controls the particle's transport properties. Thus the
model captures basic aspects of more complicated Hubbard or Holstein
Hamiltonians. Then  it is a legitimate question to ask whether the 
interaction with the background may even drive a MIT.   

Consider the Hamiltonian~\cite{Ed06} 
\begin{eqnarray}
 H= H_b-\lambda\sum_i(b_i^{\dagger}+b_i)
              +\omega_0\sum_i b_i^{\dagger}b_i+\frac{N\lambda^2}{\omega_0},
\label{model}
\end{eqnarray}
where
$ H_b=-t_b\sum_{\langle i, j \rangle} f_j^{\dagger}f_{i}
  (b_i^{\dagger}+b_j)$
describes a boson-affected nearest-neighbor hopping ($\propto t_b$) of 
spinless fermionic particles ($f_i^{(\dagger)}$). 
A fermion emits (or absorbs) a local boson $b_j^{\dagger}$ 
($b_i^{}$) every time it hops between lattice sites $i,j$. 
This way the particle creates
local distortions of certain energy 
in the background. In the case of an antiferromagnetic spin 
background the distortions correspond 
to local spin deviations (cf.\ the motion of a hole in the $t$-$J$ model
~\cite{Tr88}). If the background medium is a deformable lattice    
they are basically lattice fluctuations (phonons). Other situations
such as doped CDWs or exciton transport in molecular aggregates
might be envisaged. In any case the distortions of the background 
can be parameterized as bosons~\cite{Ed06,MH91a}. 
The distortions are able to relax 
(compare $\lambda$ with $J_\perp$ in the $t$-$J$ model),
which is described by the second term in~(\ref{model}).  The
third term gives the energy of the bosons; the constant energy shift 
$N\lambda^2/\omega_0$ guarantees finite energy for $N \to \infty$.  
Performing the unitary transformation $b_i \mapsto b_i + \lambda / \omega_0$
eliminates the boson relaxation term in favor of a second, free-fermion 
hopping channel,  $H\to H= H_b+ H_f+\omega_0\sum_i b_i^{\dagger}b_i$,
where $H_f= - t_f \sum_{\langle i, j \rangle}  f_j^{\dagger} f_i$
with $t_f=2\lambda t_b/\omega_0$. Hereafter we focus on the 1D half-filled 
band case, i.e., fermion number $N_f/N=1/2$, and take $t_b=1$ as energy unit.  

In our model~(\ref{model}), the particles have only a charge degree of 
freedom. Then, for a tight-binding band structure and in the 
absence of disorder, the formation of a CDW is 
the only possibility for a MIT. The CDW might be induced by strong
correlations in the background, which exist for large $\omega_0$
because (i) coherent transport ($\propto t_f$) takes place on a 
strongly reduced energy scale only and (ii) incoherent transport 
is energetically costly~\cite{Ed06}. Hence, in the limit 
$\omega_0\gg 1$, an effective Hamiltonian with nearest-neighbor 
fermion repulsion results. By contrast, if the local distortions  
of the background relax readily (i.e., $\lambda\gg 1$) and/or the 
energy of the bosons is small (i.e., $\omega_0\ll 1$), the free hopping channel
can act efficiently against any correlation-induced charge ordering.

Evidence for a MIT comes from a very recent exact diagonalization (ED) study
of~(\ref{model}): Calculating the wave-vector-resolved photoemission and 
inverse photoemission spectra the opening of a single-particle excitation gap 
has been observed at $K_F=\pm \pi/2$ as $\lambda$ decreases at 
relatively large $\omega_0=2$~\cite{WFAE08}. Of course, dealing with 
lattices up to 16 sites, this does not unambiguously  prove the existence 
of a true phase transition which may occur in the thermodynamic limit
$N\to \infty$ only. 

In this work,  we carry out the first 
large-scale DMRG investigation of the two-channel 
transport model~(\ref{model}). In combination with a  
finite-size scaling analysis this allows us to map out 
the ground-state phase diagram for the 1D half-filled
band case and to characterize the different phases involved. 
The DMRG is one of the most powerful 
and accurate numerical techniques for studying 1D fermionic many-body 
systems~\cite{Wh92}. It can be easily generalized to treat systems
including bosons. Within the pseudo-site approach an exact mapping 
of a boson site, containing $2^{n_b}$ states, to $n_b$ pseudo-sites
is performed~\cite{JW98b}. Here we take into account up 
to $n_b=4$ pseudo-sites, so that the $n_b$-th local boson density
is always smaller than $10^{-8}$.
In addition, we keep $m=1200$ to $2000$ density-matrix eigenstates
and extrapolate all quantities to the $m\to \infty$ limit. 
To test our DMRG implementation we compared data obtained for small systems
with previous ED results~\cite{WFAE08} and got very good 
agreement: The relative error of the ground-state energy 
$|E_{\rm ED}-E_{\rm DMRG}|$ was always smaller than 
$10^{-7}$ (for all $\lambda$ at $\omega_0=2$);
the discarded weight was smaller than $5\times 10^{-8}$.  

\begin{figure}[t]
\includegraphics[width=\linewidth]{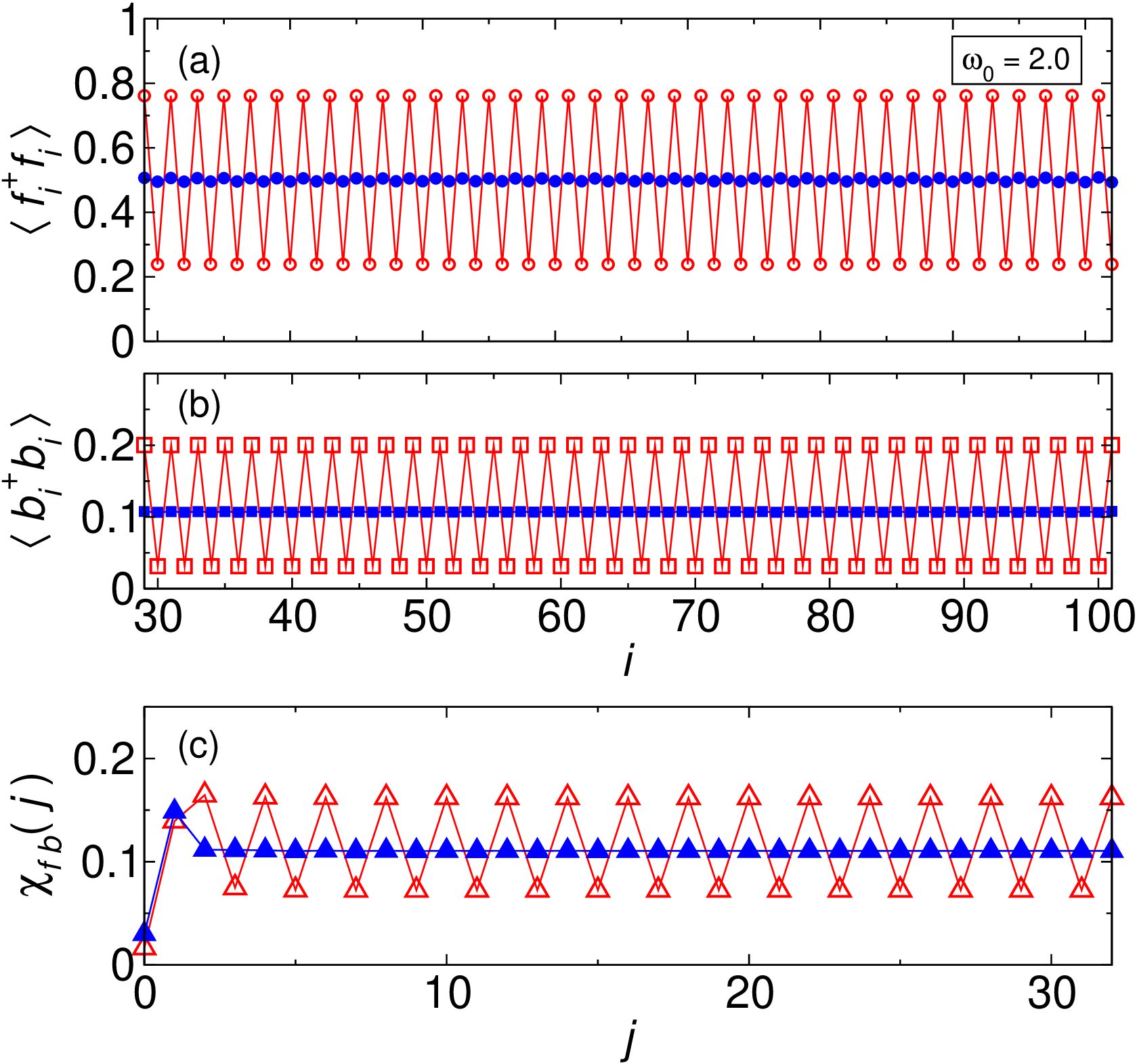}
\caption{(color online) Local densities of fermions $\langle  f_i^{\dagger} f_i \rangle$ (a) and bosons $\langle  b_i^{\dagger} b_i \rangle$ 
(b) for a 128-site system with OBC. Open symbols are 
for $\lambda=0.01$ (CDW regime), filled ones for $\lambda=0.5$ 
(metallic regime). The fermion-boson correlation function $\chi_{fb}(j)$ 
is given in panel (c) for a 64-site system with APBC
(here the discarded weight is $1.4\times 10^{-10}$
$(7.9\times 10^{-10}$) for $\lambda=0.01$ ($\lambda=0.50$)). In all cases 
$\omega_0=2.0$.
}
\label{fig1}
\end{figure}

As indicated by small cluster EDs~\cite{WFAE08}, 
at $\omega_0=2.0$, where fermions and bosons are strongly 
correlated for small $\lambda$, a MIT might occur in the range 
of $0.01 < \lambda <  5$. This is confirmed by DMRG for much 
larger systems: Figs.~\ref{fig1} (a) and (b), showing the variation of 
the local densities of fermions $\langle  f_i^{\dagger} f_i \rangle$ 
and bosons $\langle b_i^{\dagger} b_i^{} \rangle$, respectively, 
point towards the existence of a homogeneous state (CDW state)  
for rather large (small) $\lambda$. Using open boundary conditions
(OBC), the system is obviously not translation invariant,
i.e., the local density is inhomogeneous in any case.  
In the CDW phase, there are two degenerate ground states. 
Within an OBC DMRG calculation, one of these ground states is
picked out by initializing the DMRG algorithm, so that the 
CDW state is directly observable in the local density. 
In the metallic regime, on the other hand, the open boundaries 
reveal (strong) Friedel oscillations, which will be algebraically 
reduced, however, as we move towards the interior. 
Thus for large enough system sizes, within the central part of the chain, 
the local density becomes constant (see filled symbols 
in Figs.~\ref{fig1} (a) and (b))~\cite{comment1}. 
   
The CDW structure of the insulating state  shows up also 
in the fermion-boson correlation function 
\begin{eqnarray}
 \chi_{fb}(j)=\frac{1}{N_f}\sum_i
    \langle f^\dagger_{i}f^{}_{i}b^{\dagger}_{i+j}b_{i+j}^{}\rangle.
\end{eqnarray}
Calculating $\chi_{fb}(j)$ at $\omega_0=2.0$ for $N=64$ 
with anti-periodic boundary conditions (APBC)~\cite{comment2}, 
we find a distinctive alternation for $\lambda=0.01$ 
and a constant value away from the ``central site'' for 
$\lambda=0.50$ which again supports the MIT scenario
(see Fig.~\ref{fig1} (c)). Note that in the 
latter case there is still a large boson density at the particle's 
nearest-neighbor site, locally enhancing the mobility of the carrier. 

\begin{figure}[t]
\includegraphics[width=\linewidth]{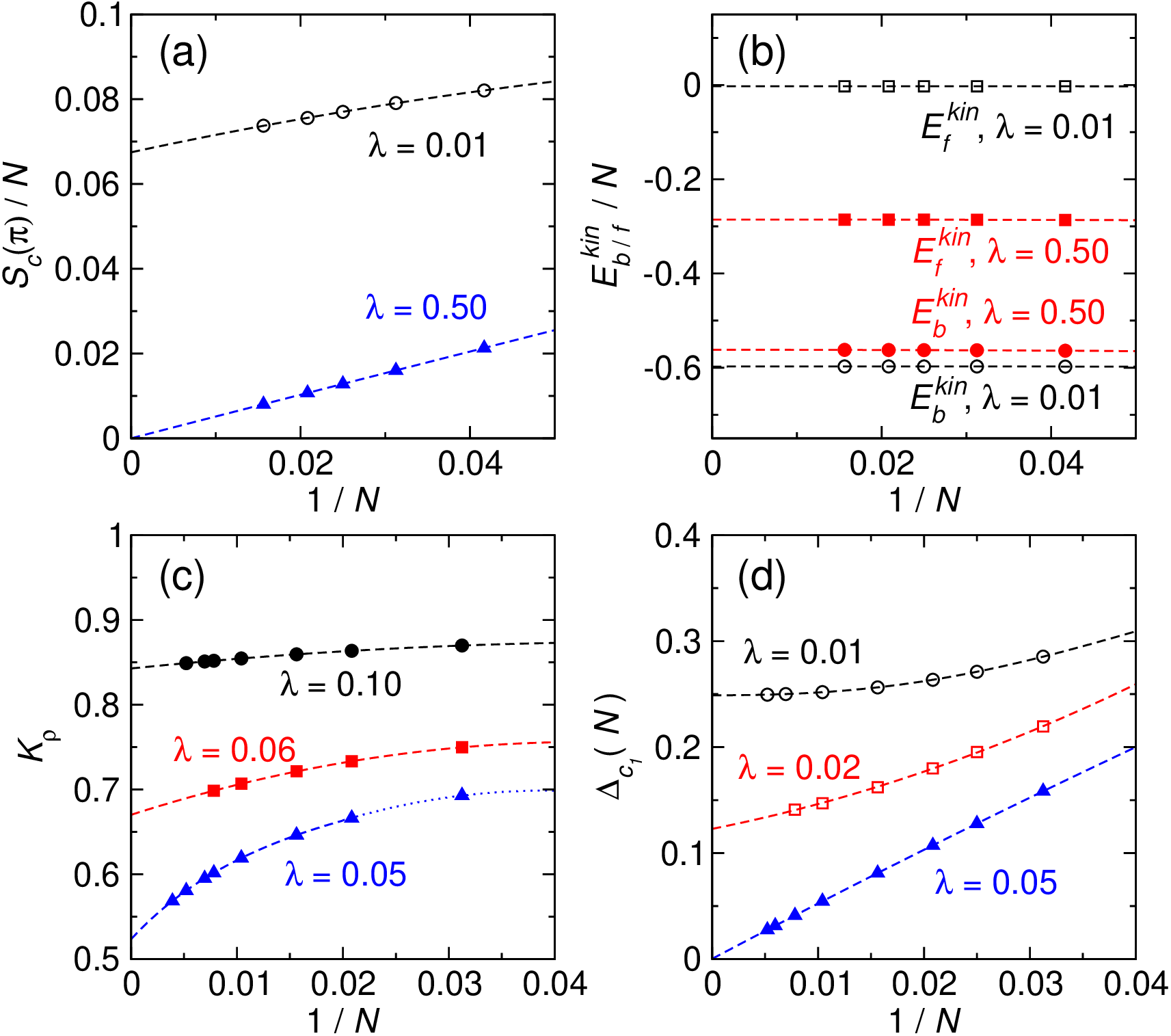}
\caption{(color online) Finite-size scaling of several physical 
quantities: (a) static charge structure factor $S_c(q)$ at $q=\pi$,  
(b) kinetic energy parts $E_{b/f}^{kin}$, (c) Luttinger liquid parameter
$K_\rho$, and (d) single-particle excitation gap $\Delta_{c_1}$. 
Data obtained for $\omega_0=2$ with APBC [(a),(b)] and OBC [(c),(d)] 
applied.}
\label{fig2}
\end{figure}

Whether the pronounced CDW correlations observed for small $\lambda$
and large $\omega_0$ are signatures of true long-range order
remains an open issue yet. To answer this question,
we explore the static charge-structure factor,
\begin{eqnarray}
 S_c(q)=\frac{1}{N}\sum_{j,k}e^{iq(j-k)}
    \left\langle \left(f^\dagger_{j}f^{}_{j} -\frac{1}{2}\right) 
    \left(f^\dagger_{k}f^{}_{k}-\frac{1}{2}\right)\right\rangle,
\end{eqnarray}
where $0\leq q < 2\pi$. If $S_c(\pi)/N$ stays finite in the 
thermodynamic limit, CDW long-range order exists. 
Fig.~\ref{fig2} (a) demonstrates that this is the case
for  $\lambda=0.01$, i.e. when the distortions of the 
background relax poorly. By contrast, $\lim_{N\to\infty} S(\pi)/N=0$
for $\lambda=0.5$. This means the model (\ref{model}) undergoes a 
quantum phase transition from a metal to an insulator 
as the relaxation parameter $\lambda$ decreases at fixed
$\omega_0$.

Next we investigate the relative importance of the 
different transport mechanisms  
by calculating the kinetic energy parts 
\begin{eqnarray}
 E_{b/f}^{kin}=\langle \psi_0|H_{b/f}|\psi_0\rangle
\end{eqnarray}
($|\psi_0\rangle$ denotes the ground state). 
Fig.~\ref{fig2} (b) shows that $E_f^{kin}$ tends to zero at small 
$\lambda$, indicating the suppression of the coherent transport
channel. Note that boson-assisted transport is possible
for both small and large $\lambda$ ($E_b^{kin}$ stays close to $-0.6$,
see panel (b)), and even becomes more pronounced in the CDW 
phase.

Finally we determine the TLL charge exponent $K_\rho$ 
and the single-particle excitation gap $\Delta_{c_1}$
(which, for the model~(\ref{model}), equals the charge gap). 
$K_\rho$ is proportional to the slope of the charge structure
factor in the long-wavelength limit $q\to 0^+$ \cite{Dz95}:
\begin{eqnarray}
 K_\rho=\pi\lim_{q\to 0}\frac{S_c(q)}{q}, \ \ \ q=\frac{2\pi}{N},\ \ \
  N\to \infty.
\end{eqnarray}
From this relation we can calculate $K_\rho$ quite accurately using 
DMRG techniques. As is well-known the 1D spinless fermion 
model with the nearest-neighbor Coulomb interaction $V$ at half filling 
can be mapped onto the exactly solvable XXZ model. 
There the TLL charge exponent decreases from $K_\rho=1$, as $V$ is enhanced, 
and finally reaches 1/2 at the MIT point \cite{EGN05}. 
We expect that this holds also for 
the 1D spinless fermion transport model~(\ref{model}) at half band-filling,
even though there are only a few analytical or numerical results 
referring to this for coupled fermion-boson systems 
(for the (Hubbard-) Holstein model see~\cite{BMH98,TAA05}).  
The single-particle (charge) gap can be obtained from
\begin{eqnarray}
 \Delta_{c_1}(N)=E(N_f+1)+E(N_f-1)-2E(N_f),
\end{eqnarray}
where $E(N_f)$ and $E(N_f\pm 1)$ are the ground-state energies in the 
$N_f$- and $(N_f\pm 1)$-particle sectors, respectively, with $N_f=N/2$. 
Fig.~\ref{fig2} (c) and (d) illustrate the finite-size scaling
analysis for the TLL parameter (c) and the charge gap (d) at
$\omega_0=2.0$. Both physical quantities can be extrapolated by
performing a least-squares fit to a second-order polynomial in $1/N$.
Note that close to the MIT points we need larger system sizes, because 
a strong  finite-size dependence evolves. In this regime,
we use chains from $N=32$ to 256 sites and higher order polynomial 
functions (up to fourth order) to extrapolate the data.  
In doing so, we determine the non-universal exponents $K_\rho(\lambda) >0.5$
in the metallic TLL phase where $\Delta_{c_1}=0$, and the finite
charge gap $\Delta_{c_1}>0$ in the CDW phase (see panels (c) and (d)).

\begin{figure}[t]
\includegraphics[width=.9\linewidth]{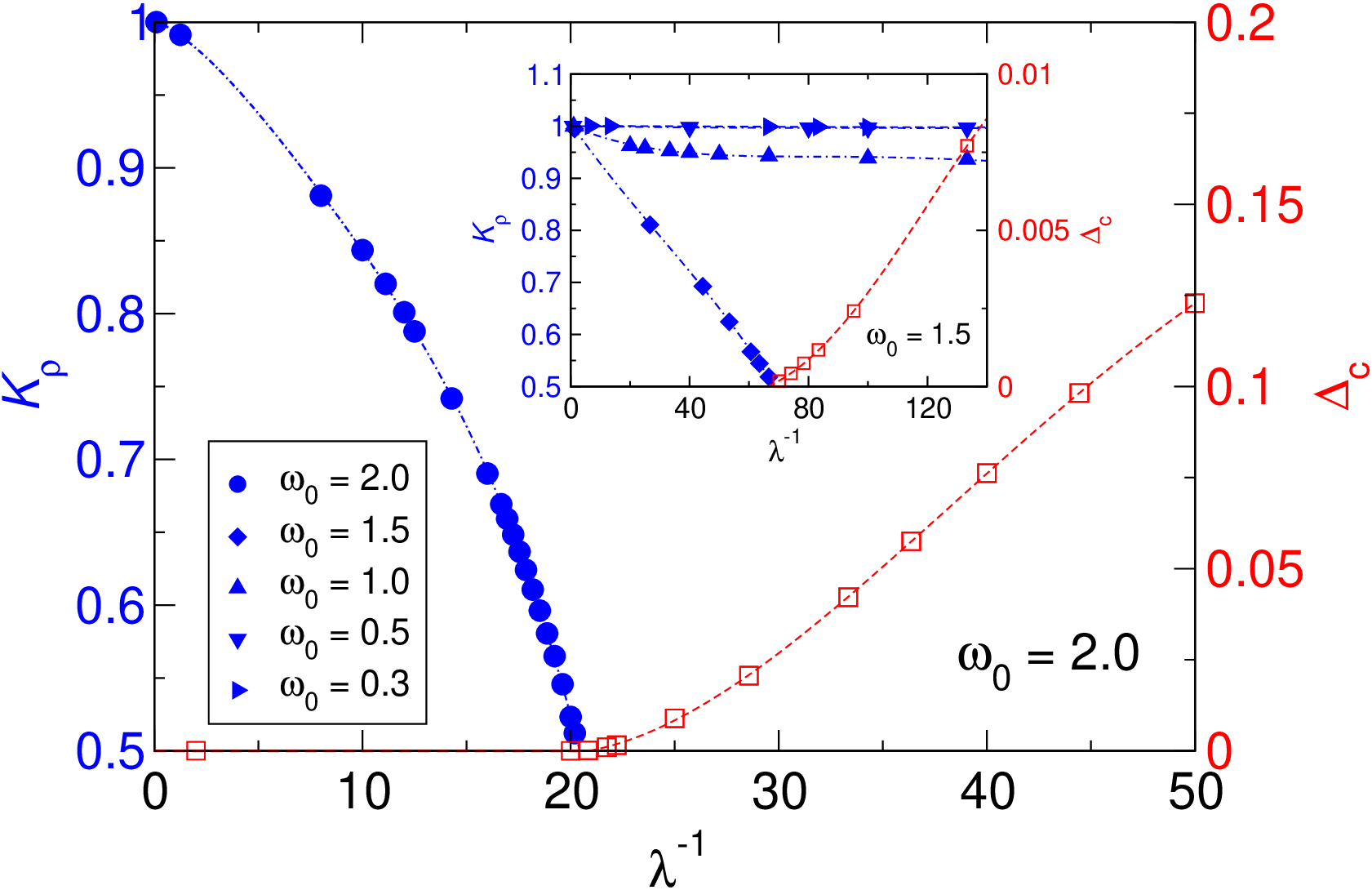}
\caption{(color online) Single-particle gap $\Delta_{c_1}$ (squares) and TLL 
parameter $K_\rho$ as a function of $\lambda^{-1}$ 
for $\omega_0=2.0$ (main panel). The inset  displays results
for smaller $\omega_0$ and shows that (i) no CDW state is found
for $\omega_0<\omega_{c}$ and (ii) $K_\rho < 1$ for all $\omega_0$,
where $K_\rho \to 1$ as $\omega_0\to 0$ and/or $\lambda^{-1}\to 0$.}
\label{fig3}
\end{figure}
In Fig.~\ref{fig3} we display the extrapolated values of 
the TLL exponent and the charge gap as a function of $\lambda$ 
at fixed $\omega_0=2.0$. Lowering $\lambda$, $K_\rho$ decreases 
from $1\to 1/2$. The point where $K_\rho=1/2$ is reached
marks the critical coupling for the MIT 
($\lambda^{-1}_c(\omega_0=2) \sim 20.4$). 
From the extrapolated DMRG data it seems that the charge gap opens 
exponentially on entering the insulating phase and afterwards 
rises almost linearly.  
This result is similar to what is observed
for the TLL-CDW transition in the anti-adiabatic strong-coupling
limit of the spinless fermion Holstein model, which there possesses
XXZ-model physics (i.e., a Kosterlitz-Thouless transition at
the spin isotropy point). But note that we find a repulsive 
particle interaction ($K_\rho \leq 1$) in the metallic phase 
for small boson frequencies as well, i.e. there is no indication 
for a pairing instability in the half-filled band case.

\begin{figure}[t]
\includegraphics[width=.9\linewidth]{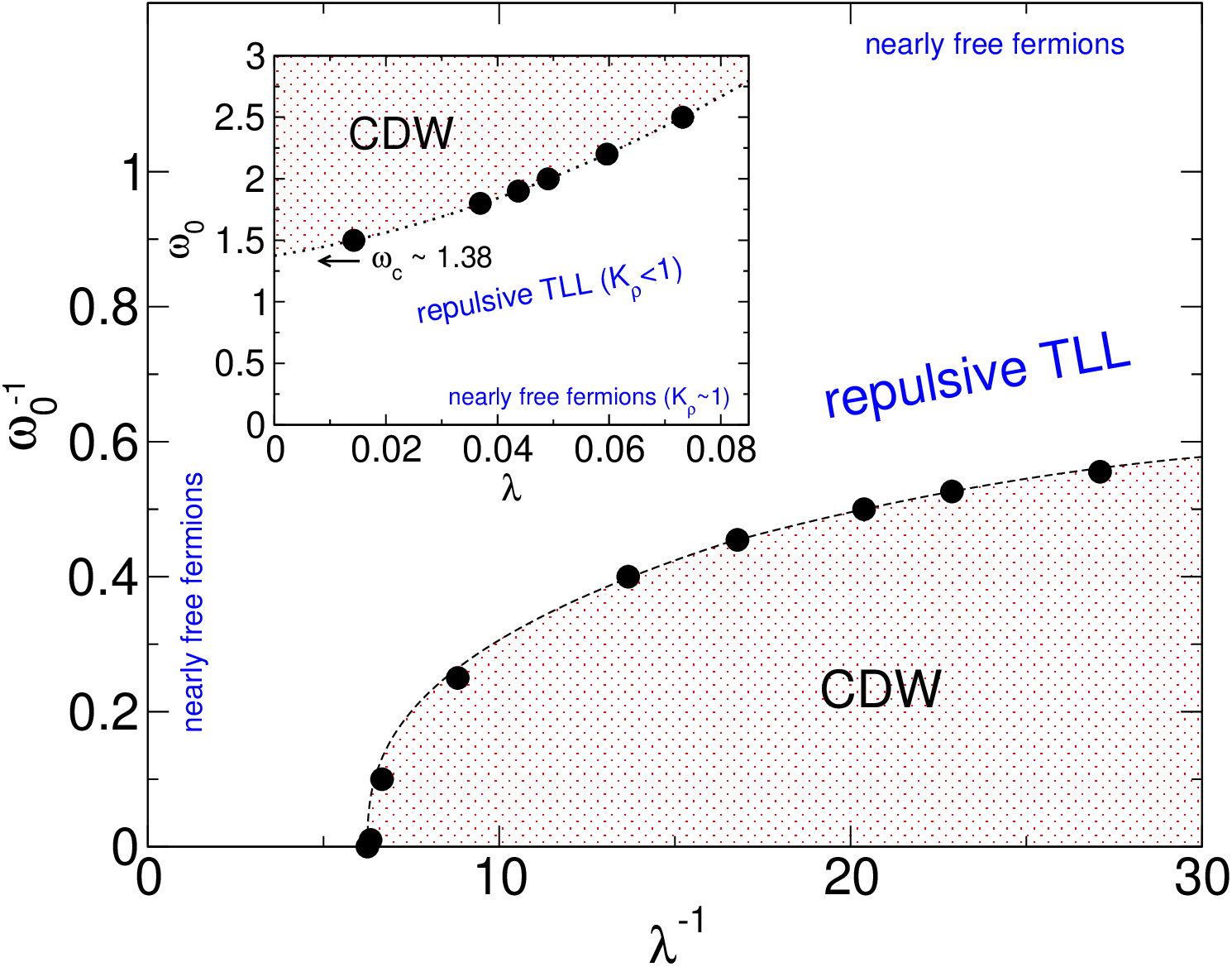}
\caption{(color online) DMRG phase diagram of the two-channel transport model
 (\ref{model}) for the 1D  half-filled band case (the dashed line 
is a guide to the eye). 
  The inset gives the phase diagram in the $\lambda$-$\omega_0$ plane. 
  The MIT  point for $\lambda=0$, $\omega_0(0)\sim 1.38$, is obtained 
  from a quadratic fit (dotted line).  
}
\label{fig4}
\end{figure}

Figure~\ref{fig4} represents the main result of our work: 
The ground-state phase diagram of the 1D half-filled fermion-boson 
model~(\ref{model}) in the $\lambda^{-1}$-$\omega_0^{-1}$ plane. 
Obviously the phase space is divided into two regimes, the metallic 
TLL phase and the insulating CDW phase with long-range order. 
We first discuss the limit of large $\omega_0$. In this regime 
background fluctuations, which are intimately connected
with any particle hop, are energetically costly. As a result 
the itinerancy of the particles is suppressed to a large extent
and charge ordering becomes favorable. Nevertheless, we
find a metallic state even for $\omega_0=\infty$ provided
that $\lambda^{-1} < \lambda_c^{-1}(\omega_0=\infty)$
(numerically we proved the TLL to exist for 
$\lambda^{-1} < \lambda_c^{-1}\simeq 6.3$ at $\omega_0=1000$).
In this case the system's ability for relaxation ($\propto \lambda$)  
is strong enough to prevent long-range charge order. 
This is reminiscent of the existence of a finite critical coupling
strength $g_c$ $(g^2=\varepsilon_p/\omega_0)$ in the 
antiadibatic limit $(\omega_0\to\infty$) of the spinless 
fermion Holstein model, where the TLL phase is realized 
for $g<g_c(\omega)$~\cite{BMH98}. In contrast to the 
TLL-CDW transition in the Holstein model, however,
the symmetry-broken CDW state is a few-boson state~\cite{WFAE08}
(i.e., not a Peierls phase with many phonons involved).  
In the opposite limit of small $\omega_0$, the background 
medium is easily disturbed by  particle motion. Therefore 
the rate of bosonic fluctuations ($\propto \omega_0^{-1}$) is high. 
Now we enter the fluctuation dominated regime~\cite{Ed06}, 
and consequently CDW order is suppressed. The inset of 
Fig.~\ref{fig4} shows that even for $\lambda=0$, i.e. if 
the explicit $\lambda$-relaxation channel is closed,    
a metallic state may exist below a finite critical energy $\omega_0(0)$.

To conclude, using an unbiased numerical (DMRG) technique, we 
proved that the very general fermion-boson transport model~(\ref{model}) 
displays a correlation-induced metal insulator transition
at half filling in 1D. The metallic phase typifies a repulsive
Luttinger liquid, while the insulating phase shows 
CDW long-range order. The phase boundary between these states
is non-trivial. It would be highly desirable to verify the
numerical results of this paper by an analytical (field theoretical
or algebraic) approach.

{\it Acknowledgments.}
The authors would like to thank A. Alvermann,  D. M. Edwards, 
E. Jeckelmann, S. Nishimoto, and  G. Wellein 
for valuable discussions. This work was supported by 
DFG through SFB 652, and the KONWIHR project HQS@HPC.
S. E. acknowledges funding by Ministerium f\"ur Bildung, Wissenschaft 
und Kultur Mecklenburg-Vorpommern, Grant No. 0770/461.01. 
\vspace*{-0.6cm}


\begin{thebibliography}{24}
\bibitem{comment2}
Note that APBC give the lowest 
energy for our spinless fermion model with $N_f=4n$ particles and 
$2 N_f$ sites.
\end{thebibliography}

\end{document}